\documentclass[preprint]{revtex4}
\def\black{}
\def\red{}

\def\blue{}

\begin{document}
\title{The disinformation problem for black holes}
\author{Sean A. Hayward}
\affiliation{Institute for Gravitational Physics and Geometry, Penn State,
University Park, PA 16802, U.S.A.\\ {\tt sean\_a\_hayward@yahoo.co.uk}}

\begin{abstract}
The supposed information paradox for black holes is based on the fundamental
misunderstanding that black holes are usefully defined by event horizons.
Understood in terms of locally defined trapping horizons, the paradox
disappears: information will escape from an evaporating black hole. According
to classical properties of trapping horizons, a general scenario is outlined
whereby a black hole evaporates completely without singularity, event horizon
or loss of energy or information.
\end{abstract}
\maketitle

According to reports, the {\em\blue information paradox\black} for black holes
has been vexing the most brilliant minds in physics for 30 years, but has
recently been resolved by Stephen Hawking using quantum gravity \cite{GR17}.
Actually, the paradox can be resolved much more simply. Logically there are no
true paradoxes, just misunderstandings, usually simple but fundamental.

\medskip\centerline{{\em\red No escape?\black}}\medskip

The classical idea of a {\em\blue black hole\black} is a region where the
gravitational field is so strong that nothing can escape, not even light. This
is a prediction of Einstein's theory of gravity, General Relativity (GR), and
nowadays astronomers regularly report observations of black holes, such as the
supermassive black hole at the centre of our galaxy.

The textbook definition of a black hole is by an {\em\blue event
horizon\black} \cite{txt}, which is the boundary between regions where a
particle can or cannot escape to {\em\blue infinity\black}, an abstract
mathematical notion. Consequently its location at any given time is determined
by the entire future history of the universe. For instance, an event horizon
could be passing through you right now due to events in the future, and you
would feel nothing. This overly abstract definition appears to be the main
conceptual obstacle causing the supposed paradox; if a black hole is defined
as a region of no escape, then, well, information cannot escape.

Hawking was perhaps the main proponent of event horizons, but has recently
stated that ``a true event horizon never forms, just an apparent horizon''
\cite{GR17}. However, Hawking's definition of {\em\blue apparent
horizon\black} \cite{txt} is also not practical for defining black holes, due
to its indirect nature and dependence on how one regards space-time as sliced
into space and time. Einstein's original insight in his earlier theory of
Special Relativity was that space and time are relative, but can be combined
as {\em\blue space-time\black}.

\medskip\centerline{{\em\red Gravity traps\black}}\medskip

Imagine a roughly spherical surface, on which a flash of light is detonated
simultaneously at each point. This produces an outgoing wavefront of
increasing area, with diverging light rays, and an ingoing wavefront of
decreasing area, with converging light rays. Now imagine that the surface
encloses a massive source such as a star. Since gravity is attractive, it
pulls back the outgoing light rays so that they diverge less than otherwise.
For a massive enough source and a small enough surface, even the outgoing
light rays could be converging. This is a {\em\blue trapped surface\black}
\cite{P}.

Separating the trapped surfaces from the untrapped surfaces, one expects a
{\em\blue marginal surface\black}, where the outgoing light rays are
instantaneously parallel. The marginal surface locates the black hole at a
given time, and as time proceeds, the marginal surfaces form a {\em\blue
trapping horizon\black} \cite{bhd}. Additionally, the ingoing light rays are
converging; if they were diverging, it would define a white hole. Lastly,
there are trapped surfaces just inside the horizon, and untrapped surfaces
just outside; if it were the other way around, one would have an inner rather
than outer horizon.

These local conditions suffice as a practical definition of a black hole.
Assuming the {\em\blue Einstein equations\black} with matter of {\em\blue
positive energy density\black}, it is straightforward to prove several results
\cite{bhd} which were generally believed or conjectured properties of black
holes. In particular, the horizon is {\em\blue one-way traversable\black}, so
that one can fall into a black hole but not escape; in GR terminology, the
horizon is space-like or null. Also, the horizon area cannot decrease, either
increasing if something is falling into it, or remaining constant. Thus a
black hole generally {\em\blue grows\black}.

Using such ideas, a new paradigm for dynamical black holes is being developed,
as a more practical alternative to the textbook paradigm describing stationary
black holes and event horizons. An important recent result \cite{en} is a law
of {\em\blue energy conservation\black}, derived from the Einstein equations,
which expresses the increase in mass of a black hole in terms of the energy
densities of the infalling matter and gravitational radiation.

\medskip\centerline{{\em\red Not so black\black}}\medskip

About 30 years ago, Hawking discovered an unexpected result in Quantum Field
Theory (QFT), which is the framework which describes all known matter in the
universe, and so all fundamental physics except for gravity. He found that a
stationary black hole will radiate with a thermal spectrum at a certain
temperature, like a black body in ordinary thermodynamics. This happens because
the quantum vacuum is continually bubbling, creating and annihilating pairs of
virtual particles. Near the horizon of the black hole, a negative-energy
particle may fall into the black hole, while a positive-energy particle
escapes to infinity. The black hole has thereby lost mass and, since its mass
and area are related, consequently contracts. Left by itself, it will slowly
shrink and, since the rate increases with decreasing size, presumably
{\em\blue evaporates\black} completely. The endpoint of evaporation is not
agreed, as one would need a theory of {\em\blue quantum gravity\black}, where
there are many competing notions but little consensus.

\medskip\centerline{{\em\red Losing it?\black}}\medskip

So what is the supposed paradox? Firstly, it is known theoretically that a
stationary black hole has just two parameters, mass and angular momentum. This
is all that can be known about the black hole from outside, however it was
formed, so almost all the information about the matter which formed it has
been lost to the outside universe. This is not what concerned most people,
since a proper accounting of information would include what was inside the
black hole. However, now imagine that the black hole evaporates. The matter
and energy inside it gradually come out again as Hawking radiation, which has
just one parameter, the Hawking temperature. So it might seem that almost all
the information about the matter which originally created the black hole has
been lost. That would contradict unitarity, a principle of QFT which preserves
information and, Hawking long argued, would therefore have to be abandoned.

\medskip\centerline{{\em\red The great escape\black}}\medskip

Apart from the final stage of evaporation, the process can be described in a
{\em\blue semi-classical\black} approximation, where the radiation is quantized
but gravity is treated classically by standard GR. QFT is supposed to determine
the energy densities of the radiation, which provide the source for the
Einstein equations, determining the space-time geometry. Actually all we need
here is the fact that the ingoing Hawking radiation has {\em\blue negative
energy density\black}. Then according to the Einstein equations, the usual
rules for trapping horizons switch: the horizon {\em\blue shrinks\black} in
area and becomes {\em\blue two-way traversable\black}; in GR terminology, the
horizon is time-like. Matter and radiation will escape, carrying information.

Hawking's thermal spectrum applies only for a {\em\blue stationary\black} black
hole, ignoring the back-reaction of the radiation on the black hole. Such a
black hole is not evaporating. If one includes the back-reaction as above, then
the black hole evaporates, but there is no reason to expect a purely thermal
spectrum.

In seeking details, one runs into ambiguities in QFT, let alone the choice of
theory of quantum gravity. Hawking favours Euclidean quantum gravity and now
claims that information is preserved \cite{GR17}. However, in terms of
trapping horizons, there never was a paradox; when a black hole is being
formed by gravitational collapse of normal matter, the horizon is one-way
traversable, while if it is evaporating, the horizon becomes two-way
traversable. Energy and information will escape then.

One aspect which is not necessarily clear from a semi-classical analysis is the
{\em\blue endpoint\black} of black-hole evaporation, since there is usually a
{\em\blue singularity\black} lurking inside the black hole, where space-time
curvature becomes infinite. However, few people believe in the physical reality
of the singularity, since it is a prediction of GR beyond its domain of
validity. Many experts seem to believe that some theory of quantum gravity will
resolve the singularity somehow.

Suppose then that the centre of the black hole never becomes singular. A
regular centre is untrapped, so there must be an {\em\blue inner trapping
horizon\black} of small (presumably near-Planck) area. According to the
Einstein equations, the rules for inner rather than outer trapping horizons
again switch: under positive energy density, it is shrinking and two-way
traversable, while under negative energy density, it is growing and one-way
traversable; in that case, collapsed matter can only exit the black hole
through it. As the black hole evaporates, the shrinking outer horizon races in
towards the growing inner horizon until the two meet, marking the endpoint of
evaporation. It seems likely that the outer and inner horizons form a single
smooth trapping horizon enclosing a compact space-time region of trapped
surfaces.

On causal grounds, everything which fell into the black hole must eventually
re-emerge. This is a very general scenario, using only classical GR with
minimal input from quantum gravity, independent of details of particular
theories. There is no singularity, no event horizon and no information paradox.

\medskip\centerline{{\em\red Orthodox paradox\black}}\medskip

In the author's understanding, there never was an information paradox or
problem for black holes, but rather a {\em\blue disinformation problem\black},
that the textbook GR definition of a black hole is by an event horizon.
Understood in terms of trapping horizons, the supposed paradox disappears.

\end{document}